\documentclass[sigconf]{acmart}

\usepackage{graphicx}
\usepackage{multirow}
\usepackage{lipsum}
\usepackage{subcaption}
\usepackage{savesym}
\usepackage{listings}

\usepackage[para]{footmisc}

\usepackage[show]{chato-notes}
\usepackage[normalem]{ulem}
\useunder{\uline}{\ul}{}
\usepackage[T1]{fontenc}
\usepackage[utf8]{inputenc}
\usepackage{xspace}
\usepackage{booktabs}

\usepackage{amsmath}

\usepackage{marginnote}
\newcommand{\pageenlarge}[1]{\enlargethispage{#1\baselineskip}}

\setcopyright{acmlicensed}
\copyrightyear{2025}
\acmYear{2025}
\acmDOI{10.1145/3705328.3748043}
\acmConference[RecSys '25]{19th ACM Conference on Recommender Systems}{September 22--26,
  2025}{Prague, Czech Republic}
\acmBooktitle{19th ACM Conference on Recommender Systems (RecSys ’25),
  September 22--26, 2025, Prague, Czech Republic}
\begin{document}

\title{Failure Prediction in Conversational Recommendation Systems}


\author{Maria Vlachou}
\affiliation{%
  \institution{University of Glasgow}
  \city{Glasgow}
  \country{UK}}
\email{m.vlachou.1@research.gla.ac.uk}


\renewcommand{\shortauthors}{Vlachou}

\begin{abstract}
In a Conversational Image Recommendation task, users can provide natural language feedback on a recommended image item, which leads to an improved recommendation in the next turn. While typical instantiations of this task assume that the user's target item will (eventually) be returned, this might often not be true, for example, the item the user seeks is not within the item catalogue. Failing to return a user's desired item can lead to user frustration, as the user needs to interact with the system for an increased number of turns. To mitigate this issue, in this paper, we introduce the task of \textit{Supervised Conversational Performance Prediction}, inspired by Query Performance Prediction (QPP) for predicting effectiveness in response to a search engine query. In this regard, we propose predictors for conversational performance that detect conversation failures using multi-turn semantic information contained in the embedded representations of retrieved image items. Specifically, our AutoEncoder-based predictor learns a compressed representation of top-retrieved items of the train turns and uses the classification labels to predict the evaluation turn. Our evaluation scenario addressed two \textit{recommendation scenarios}, by differentiating between {\em system failure}, where the system is unable to find the target, and {\em catalogue failure}, where the target does not exist in the item catalogue. In our experiments using the Shoes and FashionIQ Dresses datasets, we measure the accuracy of predictors for both system and catalogue failures. Our results demonstrate the promise of our proposed predictors for predicting system failures (existing evaluation scenario), while we detect a considerable decrease in predictive performance in the case of catalogue failure prediction (when inducing a missing item scenario) compared to system failures.
\end{abstract}



\maketitle

\section{Introduction}

\begin{figure}[tb]
\centering
\includegraphics[height=4cm]{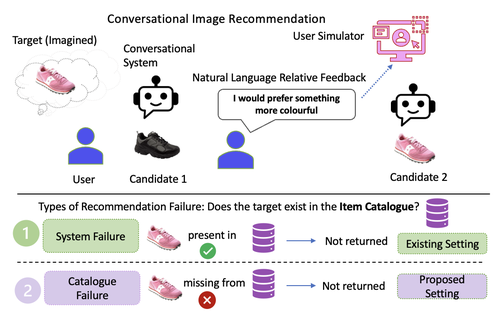}
\caption{Top: Example of an interaction in Conversational Image Recommendation. The turns develop horizontally (one CRS icon per exchange), while the simulator is used to produce the user feedback as a surrogate (same user icon in a screen). Bottom: The different cases that could lead to retrieval failures in CRS: system failure (currently implemented), and our proposed catalogue failure.}\label{fig:example_alt}\vspace{-\baselineskip}
\end{figure}

\looseness -1 Conversational Recommendation Systems (CRSs) assist users in finding items of interest by engaging in a multi-turn, goal-directed dialogue~\cite{Jannach_2021,sun2018conversational,zou2019learning}. Importantly, CRSs help with dynamic preference elicitation~\cite{christakopoulou2016towards,gao2021advances,li2018towards,sun2018conversational,zhang2018towards} by allowing users to express their preferences through natural language feedback. Specifically for online shopping, Conversational Image Recommendation~\cite{guo2018dialog,wu2020fashion,wu2021partially,wu2022multi,wu2023goal,yu2019visual} is increasingly popular, where the user sees a {\em candidate} image at each turn (as the top item of a ranking) and provides textual feedback, which describes the relative visual differences between the candidate and the user's {\em target} item. The procedure is illustrated in Figure~\ref{fig:example_alt} (top), where a user simulator is used at each turn to provide feedback on the candidate item. While in such systems the evaluation is usually based on conversation success (identification of the target item by a given turn), in a real shopping scenario, a user may not find their item(s) of interest even after a multi-turn interaction, which can lead to user frustration.

\looseness -1 For this reason, it is useful to detect {\em conversation failures}. In the existing CRSs, the user's target item is assumed to be present in the item catalogue, and failing to identify it implies the system's inability to return it. However, an item might simply be unavailable. Without knowing the exact reason for the failure, a user might keep searching for an item that is missing. However, existing research only accounts for the case where an item is present in the catalogue but not retrieved; we call this {\em system failure}. Therefore, as a first step, we need a method to detect different cases of conversational failure, including {\em catalogue failure} (Figure~\ref{fig:example_alt} (bottom), where a difference in target availability distinguishes the two failures). To this end, we are inspired by Query Performance Prediction (QPP)~\cite{carmel2010estimating}, originally used for search rankings, which predicts the effectiveness of a ranking in response to a query in the absence of relevance judgments~\cite{carmel2010estimating}. Our method predicts whether a target item (treated as a separate conversation) is found by a given rank at a certain turn.

To this aim, there are different categories of predictors to consider. First, a number of unsupervised predictors use semantic information and consider pairwise relations of top-retrieved document embeddings~\cite{arabzadeh2021query,diaz2007performance,faggioli2023geometric,vlachou2024coherence} - these are particularly applicable to image items, viewed as another form of single-representation dense retrieval. However, most of them were not designed for conversational settings~\cite{arabzadeh2021query,diaz2007performance,vlachou2024coherence}, while those used in conversational search~\cite{faggioli2023geometric,meng2023query,roitman2017enhanced} cannot generalise to a recommendation setting, where relevance judgments are not present as in IR test collections. Second, supervised BERT-based predictors~\cite{arabzadeh2021bert,datta2022pointwise,hashemi2019performance} fine-tune BERT~\cite{devlin2018bert}, i.e., use external pre-trained language models at the token level, and therefore rely on term relations; this prevents them from generalising to images. For this purpose, we introduce a method that jointly considers text and image information in a common embedding space. In parallel, to account for the lack of relevance judgments, we predict conversational performance as a classification task using the accuracy on the test set.

In short, the contributions of this paper are twofold: (i) We propose a new prediction task for CRSs and two supervised {\em conversational performance predictors} that predict failures in a recommendation dialogue. In particular, we develop an Autoencoder(AE)-based predictor~\cite{ng2011sparse,wang2014generalized} that gradually learns a low-dimension core manifold of retrieved items of multiple turns. In addition, we propose a baseline predictor that adds a shrinkage factor to existing embedding-based predictors to maintain the "important" information from the various turns. (ii) We introduce a new recommendation scenario termed {\em missing target}, and consequently, we differentiate between failures to retrieve an item ({\em system failure}) and when the target does not exist ({\em catalogue failure}). Our experiments demonstrate that our AE-based predictor is optimal for the base scenario, while our shrinkage-based predictor is the most promising for the missing target scenario. We share the source code to reproduce our proposed predictors and recommendation scenario at: \url{https://github.com/mariavlachou/failpred_missing}.

\section{Related Work}\label{sec:rw}\pageenlarge{1}
As our task is new, we need to draw inspiration from relevant work in a neighbouring task, namely query performance prediction (QPP)~\cite{carmel2010estimating}. In this regard, post-retrieval QPP predictors that employ the distribution scores of retrieved items~\cite{roitman2017enhanced,roitman2017robust,shtok2009predicting} showed promising results. Those mainly refer to the standard deviation of top-retrieved documents~\cite{shtok2009predicting} and variants of it~\cite{cummins2011improved,perez2010standard,roitman2017enhanced,tao2014query}. More importantly, coherence-based predictors use semantic relations among the top-retrieved document embeddings~\cite{arabzadeh2021query,diaz2007performance,faggioli2023geometric,vlachou2024coherence}. Those include autocorrelation (AC)~\cite{diaz2007performance}, network metrics (WAND)~\cite{arabzadeh2021query}, but also reciprocal volume (RV)~\cite{faggioli2023geometric} and A-pairRatio~\cite{vlachou2024coherence}, which additionally use the relationship of retrieved items with the query; this query-document embedding relation seems more appropriate for a conversational recommendation setting. At the same time, while the embeddings of the user feedback utterances could serve as an indicator of performance, these are produced by the user simulator during the dialogue and are limited by its capabilities, while they are quite short in length.

In a conversational setting, recent work has examined conversation continuation prediction~\cite{kongyoung2022monoqa}. More specifically for CRSs, early work on {\em Conversational Performance Prediction (CPP)}~\cite{vlachou2022performance} applied existing score-based predictors of a recommendation list to predict the rank of a target item at each turn. This unsupervised approach was limited to the case of a single or two consecutive turns. In contrast, this paper extrapolates to multi-turn prediction and develops a number of semantic supervised predictors that reflect the gradual learning of the retrieved image item representations over turns. Relevant to our approach with autoencoders is iQPP~\cite{poesina2023iqpp}, an image-based pre-retrieval prediction method that operates on the collection of images. Instead, we use the embedded representations of retrieved items from multiple turns. Unlike QPP, autoencoders are more widely used in recommender systems for different purposes, such as improving retrieval ability~\cite{spivsak2024interpretability} and personalising top-n recommendations in the cold-start problem~\cite{wu2020hybrid}.

\section{Supervised Conversational Performance Prediction}\label{sec:cpp} \pageenlarge{1}
We propose a classification task aiming to predict whether a given conversation will result in the user's target item being retrieved or not. Specifically, for a conversation $C$ consisting of $k$ turns of user critiques $c_1, \ldots c_k $, and the ranking of retrieved items $r_1, \ldots r_k$, we define a classifier $cls(X_{C,k}) \rightarrow \{0,1\}$, where $X_{C,k}$ is the feature representation for a conversation at a given turn. Our proposed approach applies constraints to the embedded representations of the retrieved image items in order to capture the important dimensions of semantic information in them. 

\subsection{Proposed Semantic Supervised Predictors}\label{ssec:cpps}
In general, we define a multi-turn feature representation of a conversation as:
\begin{equation}\label{eq:sup_cpp_1}
X^{multiple}_{C,k} = [ \gamma(\Phi_{c,1}), \ldots, \gamma(\Phi_{c,k})]
\end{equation}
where $\Phi_{c,k}$ is the embedded representation of the retrieved items at turn $k$ (the top-100 retrieved items from a single turn of the EGE model~\cite{wu2021partially} described in Section~\ref{ssec:setup}), and $\gamma$ is a function applied to $\Phi_{c,k}$ that is used as input to the $cls$. In other words, the generic representation-based multi-turn predictor $X^{multiple}_{C,k}$ (with $C$ denoting multi-turn representations) is produced from a sequence of single-turn $\gamma$ functions (with $c$ denoting single-turn representations).
Our intuition is that not all aspects of the latent representations of image items are equally informative in terms of semantics. However, unlike text retrieval models, which can detect important dimensions of embeddings at the token level, image retrieval can be seen as a form of dense retrieval with image embeddings represented by a single vector. Therefore, a predictor that captures the essence of what is being retrieved points to the reduction to the core embedding features. In this regard, Auto-Encoders~\cite{ng2011sparse,wang2014generalized} have been used in image generation and compression tasks, where they learn to compress data from the input layer into a lower dimension space, and then reconstruct it back to the original dimensions - this reconstruction is as similar as possible to the original representation. The reduced dimensional manifold represents the core dimensions of the embeddings, while $\gamma$ is the Mean Squared Error (MSE) or the reconstruction loss:
\begin{equation}\label{eq:sup_cpp_2}
\gamma(\Phi_{c,k}) = L_{rec} = (h_{w,b}(\Phi_{c,k}) - \Phi_{c,k})^2
\end{equation}
where $w$ and $b$ are the model parameters and bias, respectively, and $h()$ is the overall output function of the auto-encoder. Going further, to use the model as a classifier, we add a softmax function and a Cross-Entropy Loss as $L_{cls} = - \sum p(\Phi_{c,k})\cdot \log q(h_{w,b}(\Phi_{c,k}))$, where $p$ is the known probability distribution for each class label for an image item in the dataset, and $q$ is the approximation of the target probability distribution or the predicted probability by the model. The approximation is given by the Auto-encoder. Finally, we calculate the total loss as: $L_{total} = L_{rec} + L_{cls}$.

Another way to instantiate $\gamma$ is to use any coherence-based QPP, i.e.,:
\begin{equation}\label{eq:sup_cpp_3}
\gamma(\Phi_{c,k}) = Coherence(\Phi_{c,k}) 
\end{equation}
where $Coherence(\Phi_{c,k})$ refers to predictors that capture the pairwise relations of retrieved items such as AC~\cite{diaz2007performance} and WAND~\cite{arabzadeh2021query}, and those that also add their spatial relation to the query such as A-pairRatio~\cite{vlachou2024coherence} and Reciprocal Volume (RV)~\cite{faggioli2023geometric}. Still, the feature representations of different turns are auto-correlated, while some turns may have a higher effect in guiding the system to the target item than others. For this reason, we add an L1-based regularised variant with a shrinkage factor $\lambda$, $L_{cls} + \lambda \sum_1^n |w_i|$, which results in some of the features to be set to zero. In this way, only the "important turn" dimensions of the feature representations contribute to the prediction of the conversation label.

\subsection{Conversation Failure Evaluation}\label{ssec:scenario}
\begin{figure}[tb]
\centering
\includegraphics[width=85mm]{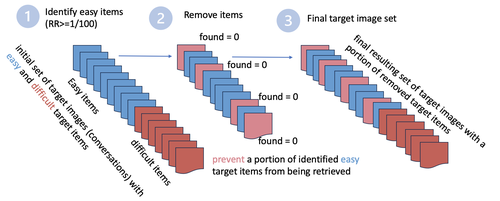}
\caption{Process of the Missing Target scenario creation. First, the set of target image items is considered, and easy items are identified based on their reciprocal rank values. Then, the Missing Target Scenario is induced by preventing some of the easy items from being removed (labeling them as not found). Finally, the categorised easy items are added back to the full set of targets.}\label{fig:missing}\vspace{-\baselineskip}
\end{figure}

For the {\em base scenario} (existing setting), we define a successful conversation as one where the target item is retrieved by rank 100 (equal to the retrieved items by the CRS) at turn $k$ (these are easy items), and a system failure otherwise (difficult items). To induce the {\em missing target} scenario, we select 30\% of easy items as follows: For {\em Conversation Failure Ground Truths}, we consider three cases for any conversation: (i) the conversation is successful, as the target item is retrieved; (ii) the conversation fails, because the system is unable to retrieve the target item based on the user’s feedback before a fixed number of turns expires (i.e., a system failure, base scenario); and (iii) the conversation fails because the system’s does not contain the target item (i.e., a catalogue failure, missing target scenario). In practice, for difficult items, which the system struggles to retrieve, there is no difference between system and catalogue failures. Therefore, to emulate catalogue failures, we sample easy items (which the system can normally retrieve successfully), and prevent them from being retrieved. When doing so, we recalculate the features. Note that both scenarios, there is the same binary classification task with labels "found" or "not found" by turn k. The difference lies in the way we induced the missing target scenario; since for difficult items, the two types of failure cannot be distinguished, we selected a portion of easy items and labeled them as not found. The missing target process is described in Figure~\ref{fig:missing} in a stepwise manner (identification of easy items, prevention from retrieval, final set of targets).

\section{Experiments}\label{sec:experiments}\pageenlarge{1}

Below, we answer the following research questions:

{\bf RQ1} How do our proposed predictors compare against other predictors: (a) in the base scenario and (b) in the missing item scenario?

{\bf RQ2} What is the impact of using: (a) single-turn predictors instead of our multi-turn predictors? (b) different rank cutoffs?

\subsection{Setup}\label{ssec:setup}
We use the Shoes~\cite{berg2010automatic,guo2018dialog} (4658 test images) and the Fashion IQ Dresses~\cite{wu2020fashion} (2454 test images) datasets, both of which contain relative critiques per candidate-target pair. For both scenarios, we use 200 sampled target items\footnote{Each target item is treated as a separate conversation.} from each dataset to induce a setting with a smaller number of per-target results, similarly to the traditional QPP evaluation setting where a smaller number of queries is used ~\cite{arabzadeh2021bert,datta2022pointwise,shtok2009predicting,shtok2010using}. The selection of images was done carefully with a QPP check to ensure varying difficulty levels, which was also reflected in the final train-train split. Following~\cite{guo2018dialog,wu2020fashion,wu2021partially,wu2022multi,yu2019visual}, we apply the Show, Attend, and Tell~\cite{xu2015show} user simulator for training an EGE CRS model~\cite{wu2021partially}, which uses historical feedback and recommendations and was found to be more effective than a supervised GRU~\cite{guo2018dialog} model. The label distribution for the missing target scenario is inevitably slightly imbalanced towards not found, as this is the condition we induced.

To implement our Auto-Encoder (AE) predictor, we use a linear activation on the first and a ReLu on the second hidden layer, and we train the model with an Adam optimizer with a learning rate of 0.01 for a total of 100 epochs. For multi-turn prediction, we use the top-100 item representations of all turns up to $k-1$ (train turns) to predict $k$ (evaluation turn). We implement the coherence-based predictors described in Section~\ref{ssec:cpps}, namely AC~\cite{diaz2007performance}, WAND~\cite{arabzadeh2021query}, RV~\cite{faggioli2023geometric}, and A-pairRatio~\cite{vlachou2024coherence}, using a scikit-learn implementation of the Random Forest classifier, while for our L1-based variant, we classify with $\lambda = 0.1$ by adapting Lasso Regression. Also, a simpler approach would be to use logistic regression for classification. Finally, we compare with a combined input from different score-based predictors (Mean, Max, and Standard deviation), a supervised approach similar to~\cite{arguello2016using,roitman2017enhanced}. To compare among the Random Forest, Logistic Regression, and L1-based variants, we select the best-performing RF predictor for each dataset. We instantiate the L1-based predictor using any of AC, WAND, RV, ApairRatio, experiment with all, and report the best-performing one. Still, for a single classifier, we use the multi-turn representation of only one of those predictors as input. For all classifiers, we use the same data splitting strategy with 70\% of the conversations for training and 30\% for testing. There are different classifiers for each turn pair (train up to turn – predict at turn).

\subsection{Results}\label{ssec:res}\pageenlarge{1}

\begin{table*}[tb]
\caption{Performance of our AE and baselines in terms of Accuracy on the evaluation turn. The best performing predictor for each CRS turn is in bold; significance (with McNemar’s test at $p < 0.05$) by comparing with the best performing baseline is indicated by (*).}\label{tab:results_base}
\resizebox{\textwidth}{!}{
\begin{tabular}{@{}lcccccccc|cccccccc@{}}
\toprule
                               & \multicolumn{8}{c|}{Shoes}                                    & \multicolumn{8}{c}{Dresses}                                   \\ \midrule
\multicolumn{1}{l|}{turn pair}      & 2,3   & 3,4   & 4,5   & 5,6   & 6,7   & 7,8   & 8,9   & 9,10  & 2,3   & 3,4   & 4,5   & 5,6   & 6,7   & 7,8   & 8,9   & 9,10  \\ \midrule
\multicolumn{17}{c}{Base Scenario} \\
\hline
\multicolumn{1}{l|}{Score-based}     & 0.80  & 0.80  & 0.83  & 0.82  & 0.82  & 0.85  & 0.83  & 0.88  & 0.60  & 0.52  & 0.52  & 0.55  & 0.53  & 0.52  & 0.50  & 0.58  \\
\multicolumn{1}{l|}{AC}        & 0.55  & 0.75  & 0.85  & 0.82  & 0.78  & 0.82  & 0.83  & 0.88  & 0.65  & 0.62  & 0.68  & 0.67  & 0.65  & 0.62  & 0.60  & 0.65  \\
\multicolumn{1}{l|}{WAND}      & 0.67  & 0.78  & 0.82  & 0.82  & 0.80  & 0.80  & 0.83  & 0.87  & 0.53  & 0.52  & 0.47  & 0.43  & 0.52  & 0.58  & 0.52  & 0.58  \\
\multicolumn{1}{l|}{A-pairRatio}       & 0.70  & 0.78  & 0.83  & 0.82  & 0.83  & \textbf{0.87}  & 0.80  & 0.88  & 0.63  & 0.52  & 0.50  & 0.47  & 0.47  & 0.50  & 0.45  & 0.52  \\
\multicolumn{1}{l|}{RV}        & 0.65  & 0.80  & 0.78  & 0.78  & 0.78  & 0.82  & 0.82  & 0.87  & 0.48  & 0.53  & 0.55  & 0.48  & 0.45  & 0.37  & 0.43  & 0.42  \\
\multicolumn{1}{l|}{LogReg}       & 0.78  & 0.82  & \textbf{0.87}  & 0.82  & 0.82  & 0.83  & 0.82  & 0.87  & 0.68  & 0.55  & 0.57  & 0.55  & 0.53  & 0.57  & 0.52  & 0.53  \\
\multicolumn{1}{l|}{AE (ours)} & \textbf{0.93}* & \textbf{0.88}* & 0.82 & 0.82 & 0.82 & 0.80 & 0.80 & 0.79 & \textbf{1.00}* & \textbf{1.00}* & \textbf{1.00}* & \textbf{1.00}* & \textbf{1.00}* & \textbf{1.00}* & \textbf{0.99}* & \textbf{0.99}* \\
\multicolumn{1}{l|}{L1-based (ours)}       & 0.78  & 0.82  & \textbf{0.87}  & 0.82  & 0.82  & 0.85  & 0.83  & 0.88  & 0.68  & 0.55  & 0.55  & 0.53  & 0.55  & 0.57  & 0.55  & 0.53  \\
\hline
\multicolumn{17}{c}{Missing Target Scenario} \\
\hline
\multicolumn{1}{l|}{Score-based}     & 0.63 & 0.58 & 0.60 & 0.52 & 0.62 & 0.62 & 0.52 & \multicolumn{1}{c|}{0.63} & 0.71 & 0.65 & 0.58 & 0.62 & 0.63 & \textbf{0.70} & 0.62 & 0.65 \\
\multicolumn{1}{l|}{AC}        & 0.43 & 0.62 & 0.62 & 0.58 & 0.52 & 0.52 & 0.55 & \multicolumn{1}{c|}{0.57} & 0.70 & 0.58 & 0.57 & 0.58 & 0.62 & 0.65 & 0.58 & 0.58 \\
\multicolumn{1}{l|}{WAND}      & 0.45 & 0.50 & 0.53 & 0.57 & 0.62 & 0.63 & 0.63 & \multicolumn{1}{c|}{0.65} & 0.68 & 0.60 & 0.47 & 0.55 & 0.57 & 0.60 & 0.57 & 0.58 \\
\multicolumn{1}{l|}{A-pairRatio}       & 0.50 & 0.62 & 0.67 & 0.72 & 0.65 & 0.63 & 0.57 & \multicolumn{1}{c|}{0.65} & 0.57 & 0.57 & 0.48 & 0.58 & 0.62 & 0.67 & 0.62 & 0.58 \\
\multicolumn{1}{l|}{RV}        & 0.55 & 0.47 & 0.48 & 0.52 & 0.55 & 0.55 & 0.60 & \multicolumn{1}{c|}{0.57} & 0.70 & 0.67 & 0.50 & 0.55 & 0.60 & 0.62 & 0.67 & \textbf{0.67} \\
\multicolumn{1}{l|}{LogReg}        & 0.53 & \textbf{0.73*} & 0.70 & \textbf{0.77*} & 0.75 & 0.67 & \textbf{0.70} & \multicolumn{1}{c|}{0.70} & 0.71 & 0.67 & 0.57 & 0.62 & 0.65 & 0.68 & 0.67 & 0.65 \\
\multicolumn{1}{l|}{AE (ours)} & 0.51 & 0.41 & 0.62 & 0.62 & 0.59 & 0.61 & 0.62 & \multicolumn{1}{c|}{0.61} & 0.71 & \textbf{0.69} & \textbf{0.68*} & \textbf{0.66} & \textbf{0.69*} & 0.68 & 0.67 & \textbf{0.67} \\
\multicolumn{1}{l|}{L1-based (ours)}        & \textbf{0.67*} & 0.70 & \textbf{0.72} & \textbf{0.77*} & \textbf{0.78*} & \textbf{0.68} & \textbf{0.70} & \multicolumn{1}{c|}{\textbf{0.72*}} & \textbf{0.72} & 0.67 & 0.57 & 0.62 & 0.65 & 0.68 & \textbf{0.68} & \textbf{0.67} \\
\bottomrule
\end{tabular}}
\end{table*}

To answer RQ1, we turn to Table~\ref{tab:results_base}, which provides the classification accuracy of our proposed predictors compared to the baselines in both scenarios. A given turn pair, i.e, "2,3", means that we train and use the features up to turn 2 and evaluate using turn 3. We first describe the results of the base scenario (RQ1(a)). For Shoes, we observe that the accuracy of the AE-based classifier is higher for early turns, and for middle and late turns, our L1-based classifier is marginally better or equal to the baseline classifiers. On the other hand, for Dresses, AE is better than the baseline classifiers and the L1-based variant, and the differences in accuracy are quite large and statistically significant from the best baseline across all turns. This indicates the utility of AE in predicting conversational failures when the target item is present; it reduces the overall difficulty of the prediction task. 

In contrast, when we examine classification performance in the Missing Target Scenario (RQ1(b)), we observe: (1) An overall reduction in accuracies when removing targets compared to the base scenario, indicating the increased difficulty of predicting catalogue failures compared to system failures. (2) A difference between datasets; A-pairRatio is overall best for Shoes, while our AE is for Dresses. In both cases, differences between predictors are marginal. (3) Accuracies in the missing target scenario are lower for Shoes than Dresses, while there is a markedly larger reduction between scenarios for Shoes. This is because Dresses is a more difficult dataset than Shoes~\cite{wu2021fashion,wu2021partially}, and since we remove targets of the same portion for both, fewer targets are found by rank 100 in Dresses, and therefore, fewer items are removed. This results in a smaller effect when changing scenario. (4) An instability of baselines, as no single predictor is optimal for all cases. (5) Accuracies are rather stable with only small differences across turns (there is no account for conversation length in the evaluation of this task; if found before turn 10, it remains found). Also, for Dresses, the accuracy is at its best and turns 2,3, and then decreases. This is not necessarily counterintuitive. Especially when the target is missing, the user (who is not aware) could keep giving uninformative feedback, which might result in less accurate predictions over time (thus indicating the usefulness of the L1 constraint). In summary, not only do all classifiers display decreased performance with missing targets, but they also perform similarly to each other.

Overall, the L1-based variant of predictors presents a noticeable increase in accuracy for both datasets compared to the corresponding RF-variant (as identified in Table~\ref{tab:results_base}, A-pairRatio is the best performing RF classifier for Shoes, and RV is for Dresses), especially in the missing target scenario. Therefore, our  L1-based classifier shows competitive performance to our AE-based predictor and can be used as a promising baseline for future evaluation on failure prediction in CRS.


\begin{table}[ht!]
\caption{Single-turn prediction results for AE in the Base Scenario using only the top-ranked item to make predictions.}
\begin{tabular}{c|c|c|c}
\hline
        & found at 1 & found at 20 & found at 100 \\ \hline
Shoes   & 0.76       & 0.57        & 0.87         \\
Dresses & 0.95       & 0.95        & 0.95         \\ \hline
\end{tabular}
\end{table}

\begin{figure}[ht!]
    \centering
     \begin{subfigure}[b]{0.45\linewidth}
     \includegraphics[width=\textwidth]{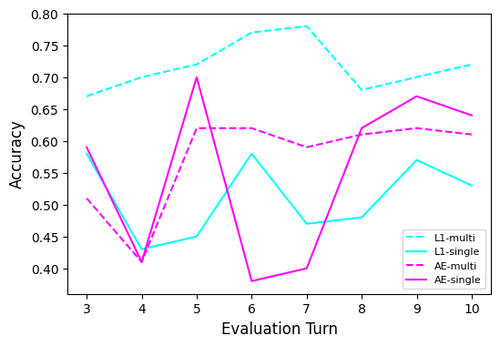}
     \caption{Shoes}
     \end{subfigure}
     \begin{subfigure}[b]{0.45\linewidth}
     \includegraphics[width=\textwidth]{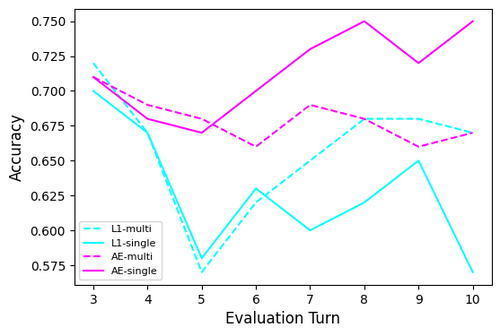}
     \caption{Dresses}
     \end{subfigure}
    \caption{\looseness -1 Comparing single-turn with multi-turn prediction for our predictors in the Missing Target Scenario.}
    \label{fig:single}
\end{figure}

To answer RQ2a), we examine Figure~\ref{fig:single} and compare for the Missing Item Scenario our proposed multi-turn evaluation setting (use features of turns up to $k-1$ to predict turn $k$) with a single-turn setting, i.e.,  use only the features of turn $k-1$ to predict turn $k$. Figure~\ref{fig:single} compares our proposed multi-turn predictors (AE and our L1-based variant of coherence-based predictors) with the corresponding single-turn predictor. The solid lines correspond to the single-turn predictors, while dashed lines are the multi-turn variants. The table demonstrates the merits of our multi-turn approach, as in most cases, single-turn results display reduced accuracy over turns.

To answer RQ2b), we examine Table 2 for the Base Scenario, which shows the sensitivity of single-turn prediction to the rank cutoff of the ground truth turn and the number of required retrieved items to use for training the AE, as shown in the case of using only the top-ranked item to make predictions. This demonstrates the usefulness of feeding the entire set of top-ranked items for training the AE in the base scenario. This also links to the surprisingly high results that we obtain in Table 1 for the Base Scenario for Dresses compared to Shoes (note that our AE operates directly on the embeddings, and does not use a per query numeric value, which adds to its usefulness).

\section{Conclusions}\label{sec:conc}\pageenlarge{1}
\looseness -1 In this paper, we proposed the new task of Supervised Conversational Performance Prediction. Inspired by QPP for search engines, we studied failure prediction in CRSs using supervised semantic multi-turn predictors. Our approach moves away from the default CRS interpretation; while the user only sees the top-ranked item, we use the embedded representations of the full set of retrieved items of the EGE model~\cite{wu2021partially} for prediction purposes, in line with traditional QPP predictors. By using these contents across turns, we show how the learned representations accumulate over time to produce multi-turn conversational performance predictors, which prove to add value on top of the corresponding predictors of a single turn. At the same time, we introduced the concept of recommendation scenarios for CRS evaluation, which predicts different types of conversational failure, namely system and catalogue failures. While our AE predictor is very effective in predicting failures due to the system's inability, our shrinkage-based multi-turn variant of coherence predictors proves to be a strong baseline that increases predictive accuracy in the case of target items not present in the catalogue (index). Our experimental results provide a first step towards supervised predictions at the conversation level in recommendation systems that use indicators capturing the core dimensions of retrieved items. As a next step, it would be useful to investigate the frequency of failures and the level of user frustration in a real shopping scenario, probably in a user study that also checks the potential of LLM-based recommendation models to generate relevance judgments equivalent to those in IR collections~\cite{faggioli2023perspectives}. We also plan to extend our predictions to different recommendation scenarios where the catalogue develops reasonably well and the user has more flexible needs that can be met with alternative relevant items~\cite{vlachou2025fashion}.

\begin{acks}
This work was supported by the UKRI Centre for Doctoral Training in Socially Intelligent Artificial Agents, Grant number EP/S02266X/1.
\end{acks}







\end{document}